\documentclass[conference]{IEEEtran}

\usepackage[
    disable,
    textwidth=.9\marginparwidth,
    colorinlistoftodos,
    prependcaption,
    linecolor=green,backgroundcolor=green!25,bordercolor=green
]{todonotes}

\makeatletter
\if@todonotes@disabled
\else \paperwidth=\dimexpr \paperwidth + 5cm\relax
\oddsidemargin=\dimexpr\oddsidemargin + 2cm\relax
\evensidemargin=\dimexpr\evensidemargin + 2cm\relax
\marginparwidth=\dimexpr \marginparwidth + 2.5cm\relax

\fi
\makeatother

\usepackage{hyperref}
\usepackage{cleveref}

\usepackage{xcolor}
\usepackage{caption}
\usepackage{graphicx}
\usepackage{minted}

\usepackage{listings}
\begin{document}

\title{Portable and Secure CI/CD for COBOL: Lessons from an Industrial Migration}

\author{
\IEEEauthorblockN{Andreas Askholm}
\IEEEauthorblockA{
Bankdata\\
Odense, Denmark\\
atw@bankdata.dk\\
ORCID: 0009-0006-8504-5321
}
\and
\IEEEauthorblockN{Kenneth Johnsen}
\IEEEauthorblockA{
Bankdata\\
Fredericia, Denmark\\
keo@bankdata.dk\\
ORCID: 0009-0002-3470-5917
}
\and
\IEEEauthorblockN{Jacopo Mauro}
\IEEEauthorblockA{
University of Southern Denmark\\
Odense, Denmark\\
mauro@imada.sdu.dk\\
ORCID: 0000-0002-5285-2868
}
}

\maketitle

\begin{abstract}
Continuous integration and delivery (CI/CD) pipelines are critical for sustaining the evolution of large software systems. In regulated industries with legacy technologies, however, pipelines themselves can become a source of technical debt. This paper presents an industrial case study of Bankdata, a cooperative IT provider for Danish banks, where a Jenkins-based COBOL CI/CD pipeline had grown fragile, slow, and tightly coupled to platform-specific logic. The original architecture relied on Groovy scripts spread across four repositories with runtime dependency installation, leading to long execution times, high maintenance costs, and vendor lock-in.

We report on the migration to a containerized architecture featuring an abstraction layer for platform logic, simplified repository structure, and a pre-built OCI-compliant image containing COBOL tools and dependencies. The new design achieved an 82\% runtime reduction. Our experience highlights lessons on abstraction, containerization, and organizational adoption, offering guidance for modernizing pipelines in legacy, high-security environments.
\end{abstract}


\section{Introduction}
\label{sec:intro}

The continuous integration and continuous delivery (CI/CD) paradigm \cite{duvall2007continuous,humble2010continuous} has become indispensable for the evolution and maintenance of large-scale software systems. In contemporary industrial practice, CI/CD pipelines form the backbone of automated testing, deployment, and delivery processes, thereby enabling organizations to sustain rapid development cycles while preserving system stability~\cite{forsgren2018accelerate}. CI/CD pipelines are critical in the banking sector, where mission-critical systems must evolve under stringent reliability, compliance, and security requirements. In this context, the efficiency, maintainability, and adaptability of CI/CD pipelines directly influence both developer productivity and the long-term sustainability of software systems.

This paper presents an industrial case study in collaboration with Bankdata~\cite{bankdata}, a Danish IT service provider that delivers core banking infrastructure to several financial institutions. Bankdata’s systems—comprised on legacy COBOL applications~\cite{iso_cobol} running on IBM mainframes—must operate with uncompromising reliability and resilience~\cite{hudson2023mainframe}. In such environments, CI/CD pipelines are not merely tools for software delivery, but strategic assets that ensure the ongoing viability of critical financial services. However, these pipelines are also subject to unique challenges: they must evolve alongside decades-old mainframe technologies, adhere to strict compliance standards, and operate within environments where vendor lock-in and long-term maintainability are persistent concerns.

The original pipeline at Bankdata was implemented using Jenkins~\cite{jenkins} and custom Groovy~\cite{groovy} scripts, defined across four repositories with tight coupling and complex runtime dependency management. This architecture, while functional, introduced significant operational friction. Dependencies were dynamically installed at runtime, leading to inconsistent environments and frequent maintenance overheads. The pipeline’s tight coupling to Jenkins increased migration barriers, locking the organization into a platform that, while powerful, imposed limitations in flexibility and adaptability. Moreover, the combination of runtime dependencies and opaque environment configurations created security concerns. These issues, when taken together, reduced developer productivity, slowed the feedback loop, and jeopardized the long-term maintainability of the software delivery process.

In this paper we present the migration of Bankdata’s legacy, Jenkins-centric pipeline to a containerized architecture. The migration pursued several goals: reducing vendor lock-in, simplifying the repository structure, improving maintainability, and ensuring compliance through pre-vetted dependencies. A key design decision was to introduce an abstraction layer that isolated platform-specific logic into a reusable package, thereby decoupling pipeline functionality from any particular CI/CD platform. At the same time, containerization was leveraged to pre-build environments that included the necessary COBOL compilers, Groovy runtime, and other command line tools. This eliminated runtime dependency installation, leading to more reproducible, secure, and performant executions.

Performance benchmarks demonstrated an 82.1\% reduction in average runtime, with pipeline execution times dropping from 724 seconds in Jenkins to 130 seconds in GitHub Actions~\cite{githubactions}. Beyond performance, the new architecture significantly improved maintainability by reducing code spread, easing onboarding, and streamlining dependency management. Security was likewise strengthened by consolidating dependencies into pre-vetted container images and by adopting stricter practices for handling secrets.

While containerization and abstraction layers are established practices, this work contributes (i) a detailed end-to-end migration blueprint in a regulated mainframe-based banking environment, (ii) an empirical characterization of migration constraints under compliance and organizational restrictions, and (iii) a combined technical and socio-organizational evaluation derived from an industrial deployment.

The rest of the paper is organized as follows. \Cref{sec:preliminaries} introduces the necessary background on CI/CD practices, Bankdata’s organizational context, and tools/technologies
 used. \Cref{sec:background} describes the pre-existing pipeline setup and outlines the objectives of the migration. \Cref{sec:migration} details the migration approach while \Cref{sec:evaluation} presents the evaluation, with both quantitative benchmarks and qualitative stakeholder feedback. \Cref{sec:discussion} discusses the lessons learned and threats to validity. Finally, \Cref{sec:conclusions} reviews related work and concludes the paper.

\section{Preliminaries}
\label{sec:preliminaries}

Before presenting the details of the migration, we provide essential background on continuous integration and delivery, the technological and organizational context of Bankdata, and the main tools and technologies underpinning our approach.

\subsection{Continuous Integration and Delivery}

Continuous Integration and Continuous Delivery (CI/CD) have become foundational practices in modern software engineering. 
The original concept of continuous integration dates back to the late 1980s with early build management systems such as Infuse~\cite{DBLP:conf/compsac/KaiserPS89}. 
Over time, this evolved into full CI/CD pipelines that automate not only build and test but also deployment, 
forming what Humble and Farley~\cite{humble2010continuous} describe as the logical conclusion of continuous integration. 
In contemporary practice, CI/CD pipelines serve as the central mechanism for enabling the release of software as frequent as possible and ensuring that software changes can be validated, built, and deployed reliably.  

At their core, CI/CD pipelines orchestrate a series of automated stages. 
Following lightweight pre-commit checks, code is integrated and subjected to developer test suites, static analysis, and integration testing. 
These validations are crucial for detecting regressions and enforcing secure coding practices~\cite{bass2015devops}. 
Compilation then produces distributable artifacts, often containerized to guarantee consistent execution across heterogeneous environments. 
The delivery stage promotes artifacts to staging, where both functional and non-functional requirements—such as performance, security, and user acceptance—are assessed before deployment. 
Deployment strategies vary with context: single-instance systems may follow a simple restart model, whereas distributed services rely on rolling or blue/green upgrades to minimize downtime.  
Moreover, modern version control systems extend the range of pipeline triggers. 
Operations such as branching, merging, or submitting a pull request each correspond to different levels of pipeline execution. 
Feature branches may only reach staging, while merges into the main branch typically trigger full validation and deployment~\cite{progitbook}. 

The abundance of available CI/CD tools reflects the maturity of the DevOps ecosystem. 
Selection often depends on compatibility, vendor support, cost, and platform constraints~\cite{cncf_landscape}. 
Because organizational needs evolve, migrations between tools are frequent. 
Vendor lock-in remains a persistent concern, as pipelines tightly coupled to a single system increase future migration costs. 
Security considerations further complicate pipeline design. 
The DevSecOps paradigm emphasizes embedding security throughout the software lifecycle, 
a necessity in light of widespread supply chain attacks~\cite{supply_chain,bird2020devsecops}. 

\subsection{Bankdata and the COBOL Legacy}

Bankdata~\cite{bankdata} is a cooperative IT service provider owned by a consortium of Danish banks. 
Its responsibility extends to delivering and maintaining a broad portfolio of banking solutions, 
ranging from payment systems and credit management to digital self-service applications and infrastructure services. 
Through these systems, Bankdata indirectly serves millions of Danish citizens on a daily basis, 
making it one of the cornerstones of the country’s financial IT infrastructure.  

A distinguishing characteristic of Bankdata’s technological foundation is the reliance on 
legacy COBOL applications deployed on IBM mainframes~\cite{ibm_mainframe}. 
COBOL has been the backbone of critical banking software for decades, valued for its proven reliability, precision in handling large-scale financial data, 
and well-understood execution semantics. 
Despite its age, COBOL remains indispensable for modern banking because it ensures stability in mission-critical processes such as transaction handling and customer data management~\cite{cobolbanking}. 

At Bankdata, the mainframe and COBOL ecosystem is intertwined with strict compliance and security requirements. 
Changes to source code or deployment procedures must undergo rigorous testing, review, and auditing. 
This regulatory environment increases the complexity of evolving the system while at the same time limiting the flexibility of 
development teams to adopt new tools or methodologies without careful evaluation.  
In this setting, the CI/CD pipeline is not simply a technical tool but an essential governance mechanism that ensures software evolution 
does not jeopardize the availability and security of banking services.  

The challenge at Bankdata stems from combining the modern CI/CD practices with a legacy stack. 
While DevOps emphasises flexibility, automation, and rapid feedback, COBOL and mainframes are associated with rigidity, 
proprietary toolchains, and long release cycles. 
This tension necessitates custom engineering solutions where CI/CD must bridge modern developer expectations with the limitations of decades-old systems.  

\subsection{Languages and Technologies}

This subsection briefly introduces the main languages and technologies referenced in this work.

\textbf{Java.} Java~\cite{java} is a widely used, general-purpose, object-oriented programming language. It provides a robust runtime environment through the Java Virtual Machine (JVM) and is commonly adopted for enterprise applications, build tools, and scripting environments.

\textbf{Groovy.} Groovy~\cite{groovy} is a dynamic, optionally typed programming language that runs on the JVM. It offers concise syntax, seamless interoperability with Java, and is frequently used for scripting and defining build or pipeline logic.

\textbf{Gradle}. Gradle~\cite{gradle} is a modern build automation for compiling code, managing dependencies, and orchestrating complex build workflows. It supports both declarative and imperative build styles and offers high extensibility through plugins. Its incremental build mechanism and strong integration with the JVM ecosystem make it well suited for automating tasks within the pipeline while ensuring reproducible and maintainable build processes.

\textbf{COBOL.} COBOL (Common Business-Oriented Language)~\cite{iso_cobol} is one of the earliest high-level programming languages, originally designed for business and administrative systems. It remains prevalent in the financial sector due to its reliability, precision in handling large volumes of data, and long-established execution semantics.

\textbf{GnuCOBOL.} GnuCOBOL~\cite{gnucobol} is a free, open-source COBOL compiler. It translates COBOL source code into C, which can then be compiled with a native C compiler. This approach enables COBOL programs to run on a wide range of modern platforms beyond mainframes.

\textbf{COBOL Expander.} COBOL Expander is an in-house tool developed by Bankdata to preprocess COBOL files
for testing. The tool begins by stripping internal comments from source files while simultaneously identifying and cataloging all referenced COBOL copybooks. These copybooks
are retrieved from the mainframe and embedded into the source files creating expanded versions ready for testing.

\textbf{COBOL Check.} COBOL Check~\cite{cobolcheck} is an open-source unit testing framework for COBOL. Inspired by modern testing practices, it allows developers to define test cases that validate COBOL program behavior in a structured, automated fashion.

\textbf{Zowe CLI.} The Zowe Command Line Interface is part of the Zowe open-source framework~\cite{zowe} and provides a standardized way to interact with IBM z/OS mainframes. It exposes commands for file transfers, job submission, dataset management, and system queries, offering a bridge between mainframe environments and modern development tools.

\textbf{Node.js and npm.} Node.js~\cite{nodejs} is a JavaScript runtime built on the V8 engine, enabling server-side and scripting applications in JavaScript. Its accompanying package manager, npm~\cite{npm}, provides reusable libraries and tools, simplifying software installation and dependency management.

\textbf{Jenkins.} Jenkins~\cite{jenkins} is an open-source automation server used for building, testing, and deploying software. It provides a plugin-rich ecosystem and pipeline-as-code functionality, supporting flexible CI/CD workflows across diverse environments.

\textbf{GitHub Actions.} GitHub Actions~\cite{githubactions} is a workflow automation service integrated into GitHub. It enables the definition and execution of CI/CD pipelines through YAML-based configuration files, supporting automation of testing, building, and deployment directly from repositories.

\section{Pre-existing Setup and Objectives}
\label{sec:background}

The initial pipeline architecture employed by Bankdata’s COBOL development teams relied on a Jenkins-based system for mainframe application deployment. The workflow was implemented using Groovy scripts tightly coupled with Jenkins-specific functionality. The implementation was distributed across four interdependent repositories, each fulfilling distinct roles, resulting in a fragmented and complex ecosystem:

\begin{itemize}
\item {Template Repository.} This repository provided standard container templates, which were compiled into a single manifest during the pipeline setup phase. The resulting containers were invoked at different execution stages.
\item {Worker Repository.} Contained the scripts defining the pipeline’s execution logic, including code retrieval, testing, compilation, and deployment. These scripts primarily orchestrated workflow sequencing while delegating specific operations to the tools repository.
\item {Tools Repository.} Hosted reusable functionality, including libraries for JSON parsing, string manipulation, and pipeline-specific Groovy scripts. Many functions executed conditional shell commands, formatted their outputs, and returned processed results to the pipeline.
\item {Configuration Repository.} Supplied templated configuration settings, including tool versions and source repository parameters. New projects were typically initiated by cloning this repository, which enforced strict directory structures to ensure correct test and source code identification.
\end{itemize}

Jenkins executes pipeline code through the Groovy Continuation-Passing Style (CPS) transformation. In this model, functions are rewritten to include explicit continuation arguments, enabling execution state to be serialized and persisted without relying on the call stack. Jenkins employs the \verb|NonCPS| annotation to exclude specific functions from this transformation, ensuring compatibility with standard Groovy code. This approach, however, introduces fragility since CPS-transformed functions can be paused and resumed reliably, whereas interactions with non-CPS code may yield execution exceptions due to the absence of continuations. Such failures safeguard against inconsistent execution states but highlight the risks inherent in mixing CPS and non-CPS paradigms.

Although operational, the baseline pipeline revealed the following structural shortcomings:
\begin{enumerate}
    \item {Complexity.} Fragmentation across multiple repositories, the presence of deprecated artifacts, and legacy code obscured critical execution paths, increased cognitive load for maintainers, and complicated modification workflows.  
    \item {Inefficient Dependency Handling.} The COBOL Expander dependency was installed during pipeline runtime, incurring unnecessary delays.
    \item {Testing.} The original pipeline lacked systematic testing mechanisms, leaving defects undetected and increasing operational risk.
    \item {Vendor Lock-in.} Heavy reliance on Jenkins introduced fragility against API changes and created migration barriers.
\end{enumerate}

\subsection{Objectives}
The migration effort aimed at addressing the previously mentioned issues. The migration was therefore guided by the following objectives.

\emph{Reduced migration costs.} Moving the pipeline logic to new platforms should require only minimal re-engineering effort.  
    The design must support this by isolating platform-specific logic and making the core functionality reusable.  

\emph{Maintainability and compliance.} The new architecture must reduce code duplication, simplify repository structure, 
    and allow easier auditing of dependencies to comply with financial-sector security requirements.  

\emph{Performance improvement.} Execution time of pipeline runs should be significantly reduced 
    to improve feedback cycles for developers and increase productivity.  

\emph{Cross-platform compatibility.} The pipeline should no longer be tightly coupled to Jenkins. 
    Instead, it must be portable across multiple CI/CD platforms (e.g., GitHub Actions, GitLab CI) thereby reducing vendor lock-in and enabling future adaptability.  

The migration, however, had to be performed under strict constraints to safeguard business continuity and maintain developer productivity. In particular no changes were allowed to the workflow of the pipeline and the business logic of the COBOL applications themselves. While obsolete or unused dependencies should be removed, 
    those that ensured stability and compatibility with mainframe tools had to be preserved. Moreover, the migration should not impose excessive disruption on existing development practices, 
    ensuring that the pipeline remained accessible to developers with varying levels of expertise in CI/CD.

\section{Migration Approach}
\label{sec:migration}

This section details the migration strategy from a Jenkins/Groovy mainframe-targeted pipeline to a containerized architecture.
We start by providing a detailed discussion of the core refactoring tasks of decoupling the pipeline from Jenkins to then focus on the dependencies and minimizing the container images. Finally, we describe the deployment of the refactored pipeline on GitHub Actions to demonstrate the portability of the approach.

\subsection{Decoupling from Jenkins}
The initial and most critical step in the migration was the decoupling of Jenkins-specific functions from the Groovy pipeline logic. Prior to refactoring, the pipeline code relied heavily on Jenkins-provided functions, which directly embedded platform-specific functionality into the workflow. This tight coupling significantly complicated maintenance, as even minor changes in Jenkins APIs necessitated widespread modifications across the codebase.

To address this, the first task was to systematically identify all Jenkins-provided functions actively used in the pipeline, together with a catalog of the functionalities they supported. On the basis of this inventory, a replacement strategy was developed. We aim to avoid Jenkins-specific calls to be eliminated in an ad hoc manner, but instead to abstract them into a structured interface that would serve as a translation layer between the pipeline code and the underlying platform.

This translation layer was implemented as a Groovy package that redefined each of the identified Jenkins functions. The pipeline code was modified to invoke these translated functions exclusively, rather than directly calling Jenkins APIs. Every point of contact between the pipeline and Jenkins was redirected through this single, centralized package. By adopting this indirection, future migrations to new build systems would require modifications within the translation layer, instead of invasive changes scattered throughout the entire codebase.

\begin{figure*}[t]
        \begin{minipage}[b]{0.45\textwidth}
        \centering
        \includegraphics[width=\textwidth]{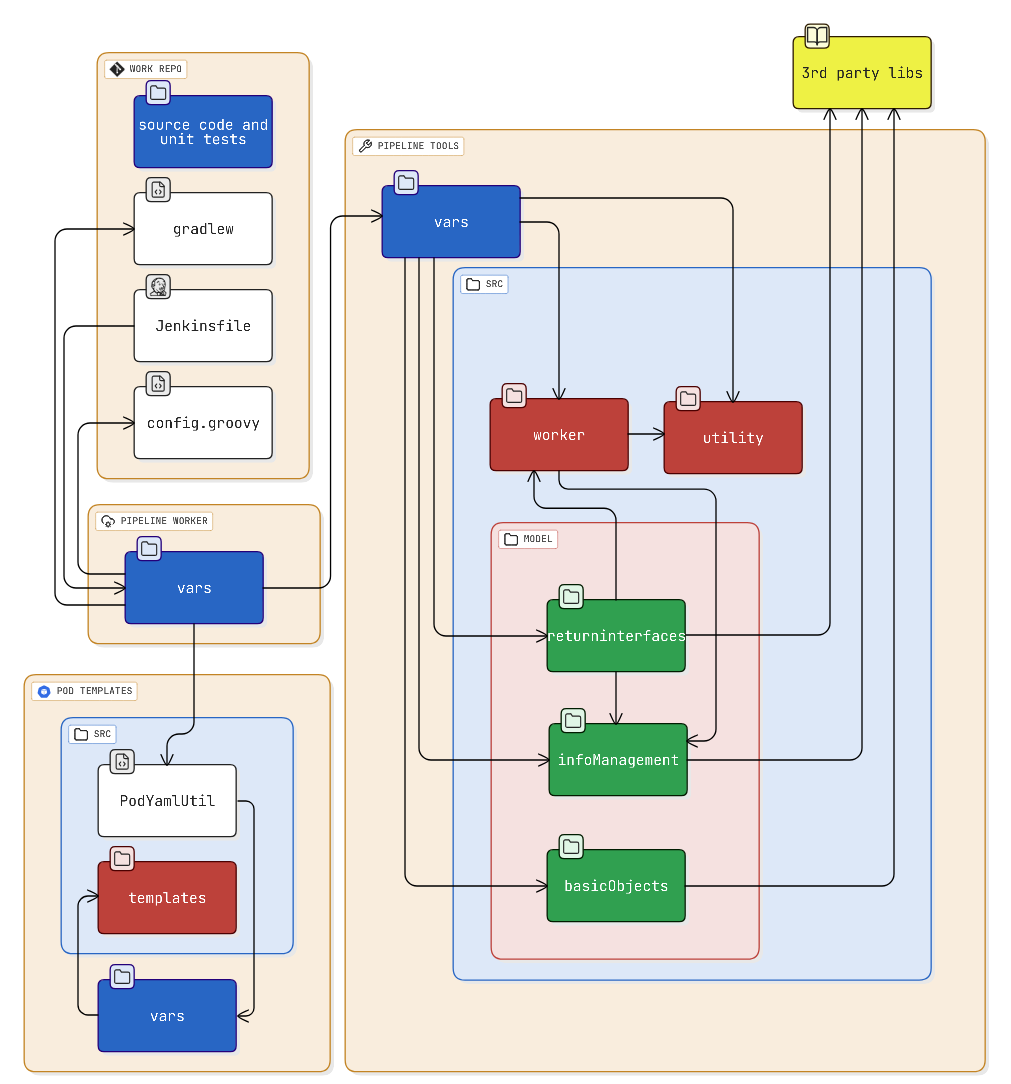}
    \end{minipage}
        \begin{minipage}[b]{0.45\textwidth}
        \centering
        \includegraphics[width=\textwidth]{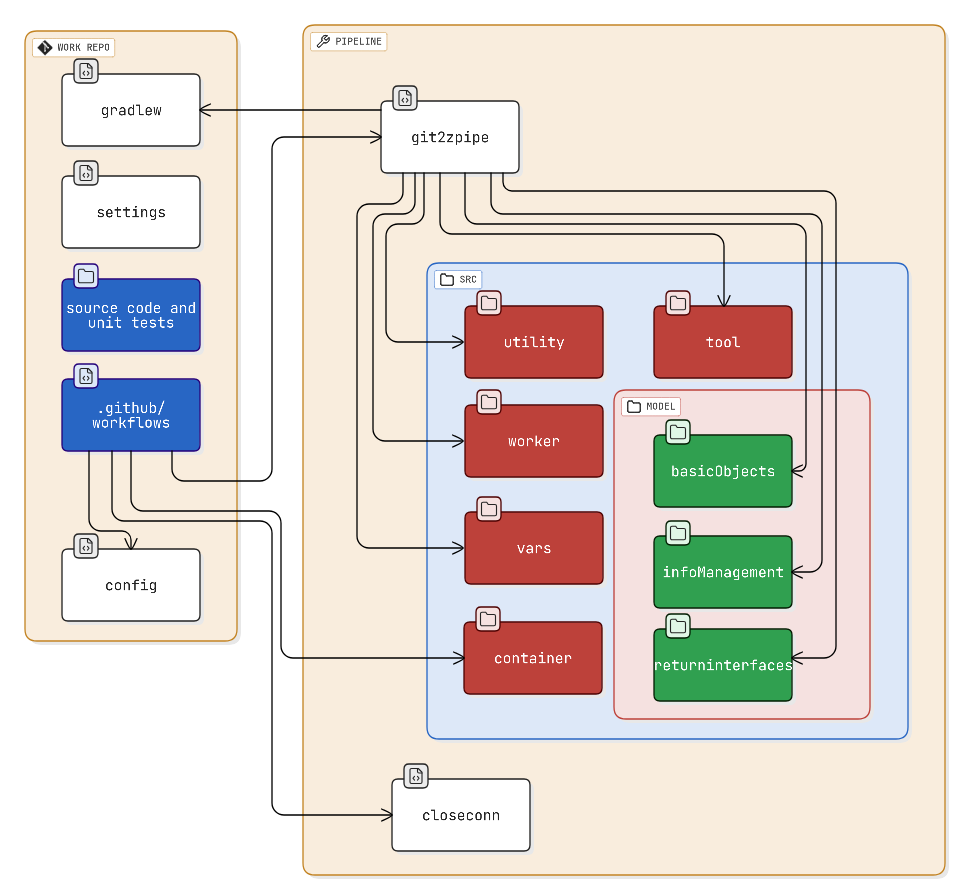}
    \end{minipage}
\caption{Visual comparison between the Groovy files constituting the pipeline before (left) and after (right) 
    the migration. Colors indicate directory depth (orange = repository root, blue = sub-dir, red = sub-sub-dir, green = deeper levels; yellow highlights Jenkins specific external libraries. Arrows represent calls or dependencies between files.
        }
\label{fig:complexity_reduction}
\end{figure*}

To give a glimpse of how the migration reshaped the overall structure, Figure~\ref{fig:complexity_reduction} compares the pipeline architecture before and after the redesign.
The left 
image depicts the original Jenkins-based setup, which was fragmented across 
multiple repositories and tightly coupled to platform-specific functionality. 
The new design (right) reduces complexity by consolidating multiple fragmented and highly dependent repositories into a single, unified one. This consolidation not only streamlines interactions but also removes the cognitive overhead of navigating between different repositories to locate relevant files or functionality. As a result, developers face fewer barriers when onboarding or making modifications, and the pipeline becomes easier to maintain and evolve over time.

\subsubsection{Managing the Environment}

One of the most pervasive challenges in disentangling Jenkins from the pipeline code lay in the area of environment management. Jenkins pipelines define execution context through a layered hierarchy of constructs. At the highest level, declarative pipelines are organized using the \verb|stages| block, which acts as the overarching container for workflow elements. Within this structure, individual \verb|stage| directives encapsulate logically related tasks, while \verb|steps| specify the concrete actions to execute. Furthermore, the \verb|script| directive provides a mechanism to embed scripted pipeline code within an otherwise declarative framework, offering additional flexibility for complex workflows.  

The example in \Cref{code:environment} illustrates these principles through the
“Mainframe Continuous Integration” stage. Here, scripted blocks such as
\verb|unit tests| provide fine-grained operational visibility within the Jenkins
interface, enabling developers to monitor progress and debug failures.

\begin{listing}[t]
  \centering
  \caption{Use of \texttt{stage}, \texttt{steps}, \texttt{container}, and \texttt{withEnv} in a Jenkins pipeline.}
  \label{code:environment}
\begin{verbatim}
stages {
  // ...
  stage('Mainframe continuous integration') {
    steps {
      script {
        // ...
        stage('unit tests') {
          // ...
          container('cobolcheck') {
            withEnv(["PATH=$PATH"]) {
              rc = runUnitTest(file)
            }
          }
        }
      }
    }
  }
}
\end{verbatim}
\end{listing}

Jenkins also provides specialized constructs for handling execution contexts and environment variables. The \verb|container| directive ensures that specified commands are executed within a given container, thereby controlling runtime dependencies. Similarly, \verb|withEnv| manages environment variable propagation across container boundaries, ensuring consistency in values such as the \verb|PATH| variable. In addition, Jenkins maintains pipeline state metadata via the global \verb|currentBuild| object, which exposes details such as commit references, repository identifiers, and build status. Together, these constructs offer a highly integrated but tightly coupled model of environment and state management.  

The migration strategy aimed to remove this dependency while preserving equivalent functionality. This was accomplished through four coordinated measures:  

\begin{enumerate}
    \item \textbf{Structural decomposition.} Monolithic pipeline blocks were refactored into functions performing a single task and then orchestrated from a central controller script. This reduced complexity and improved readability while enabling platform-agnostic reuse.  
    \item \textbf{\texttt{currentBuild} avoidance.} Information that Jenkins previously exposed through its internal objects was instead retrieved when possible from environment variables or directly through Git commands. This avoided reliance on Jenkins’ runtime metadata.  
    \item \textbf{Progress visualization.} The role of Jenkins’ stage directives in providing execution visibility was replaced with strategically placed print statements, ensuring equivalent transparency while remaining independent of platform-specific features.  
    \item \textbf{Unified containerization.} By standardizing the entire pipeline to run within a single container image, the need for Jenkins’ \verb|container| and \verb|withEnv| directives was eliminated entirely.  
\end{enumerate}

This multi-pronged approach not only preserved the essential semantics of the Jenkins environment model but also significantly simplified the execution model. By consolidating environment management into explicit, tool-independent constructs, the pipeline gained in both transparency and portability, without sacrificing functionality.

\subsubsection{Executing Shell Scripts}\label{sec:exe}

Shell execution was another critical point of Jenkins dependency. In the original setup, Jenkins’ \verb|sh| function provided a straightforward interface for executing shell commands, while the \verb|withCredentials| directive ensured secure handling of sensitive data such as mainframe login information. Credentials were injected as temporary environment variables and masked in logs, reducing the risk of exposure.  

To achieve portability, these mechanisms were replaced by custom Groovy functions. A generic \verb|shell| function was implemented to execute system commands using \verb|ProcessBuilder|, capture outputs, and return results to the pipeline. This provided equivalent functionality to Jenkins’ \verb|sh| function while avoiding platform coupling. Error handling was delegated to the caller, providing greater flexibility for adapting failure policies.  

A complementary \verb|credentials| function was introduced to handle secrets. It replaced placeholders in command strings with environment variables, providing an analogue to Jenkins’ \verb|withCredentials|. Although this approach does not natively mask secrets in logs, it was combined with log-filtering strategies to ensure compliance with security requirements.  

The resulting abstraction allowed sensitive mainframe interactions—such as authentication through Zowe CLI—to be executed in a secure and platform-agnostic manner.  

\subsubsection{Status Reporting}\label{sec:statusreport}

In the Jenkins implementation, status reporting relied on a suite of specialized functions. The \verb|echo| function provided log output, \verb|error| simultaneously logged messages and terminated execution, and \verb|unstable| flagged builds as partially successful. Additionally, \verb|catchError| blocks enabled error containment, allowing execution to continue despite failures.  

These constructs were replaced by standard Groovy logging mechanisms. Informational output is handled by \verb|println|, while critical errors are directed to \verb|System.err| followed by process termination. The \verb|unstable| state was reinterpreted as a diagnostic log message, preserving informational content without introducing tool-specific semantics. Finally, error handling was refactored to use traditional try-catch blocks, which provide equivalent control flow in a portable manner.  

This redesign preserved the functionality of status reporting while eliminating reliance on Jenkins’ proprietary functions. The result is a simpler, more explicit logging system that remains effective across different CI/CD environments.  

\subsubsection{Filesystem and Version Control}\label{sec:fsvcs}

The pipeline also required frequent interactions with the filesystem and version control system. In Jenkins, these were facilitated through functions such as \verb|checkout|, \verb|dir|, \verb|deleteDir|, and \verb|writeFile|.  

In the migrated implementation, version control operations are delegated to the hosting CI/CD platform—in the case of GitHub Actions, the \verb|actions/checkout| module. Strict version pinning is enforced to prevent vulnerabilities arising from mutable version tags.  

Filesystem operations were re-implemented using Groovy’s native \verb|File| class, covering file I/O, directory deletion, and existence checks. This ensured equivalent functionality while avoiding Jenkins dependencies.  

By transitioning to native constructs, filesystem and version control operations became simpler, more transparent, and portable across platforms.  

\subsection{Dependencies and Containerization}

While the migration sought to minimize third-party dependencies, several tools were indispensable for functionality. Each was carefully evaluated with respect to both technical necessity and security implications. In particular, the following dependencies were retained.

\begin{itemize}
    \item \textbf{Gradle:} Used to compile and execute unit tests for the pipeline code and to prepare libraries required for COBOL testing.
    \item \textbf{Zowe CLI:} Provides the interface to IBM z/OS mainframes. Its privileged access makes it indispensable but also introduces significant security considerations. Rigorous configuration management and version pinning are required.  
    \item \textbf{GnuCOBOL:} Enables COBOL programs to be compiled into C code and executed on Linux systems, forming the basis for off-platform unit testing.
    \item \textbf{COBOL Expander:} An internally developed tool that preprocesses COBOL files by expanding copybooks.
    \item \textbf{COBOL Check:} An open-source testing framework used to validate COBOL programs before deployment. Its source-modifying behaviour necessitates careful verification of test cases and strict dependency control.  
    \item \textbf{Node.js and npm:} Required for installing Zowe CLI and COBOL Expander.
\end{itemize}

By explicitly cataloguing and securing these dependencies, the pipeline balances functionality with compliance to stringent financial-sector security requirements.  

To ensure consistency across environments, the pipeline is executed within a custom OCI-compliant container image. This image includes all necessary tools (viz., Groovy, Zowe CLI, COBOL Expander) eliminating the need for runtime installation.  

The use of containerization brought three immediate benefits: faster feedback cycles, lower runtime latency, and reproducible execution environments. However, it also introduced challenges related to image size and build time.  

\begin{table*}[t]
\centering
\begin{tabular}{|c|c|c|c|c|}
\hline
Step & Change & Size & Change & Build Time \\
\hline
\hline
1 & Initial image based on debian-slim & 1423 MiB & 0 & 187.5 s \\
\hline
2 & Use alpine as base instead of debian-slim & 813 MiB & 610 MiB & 93.7 s \\
\hline
3 & no cache flag when installing from apk & 813 MiB & 0 MiB & 87.3 s \\
\hline
4 & Use openjdk21-jre instead of openjdk21-jre-headless & 817 MiB & +4 MiB & 94.5 s \\
\hline
5 & Reduce RUN commands and revert to headless & 813 MiB & 4 MiB & 87 s \\
\hline
6 & Explicitly delete apk cache with rm command & 783 MiB & 30 MiB & 87.5 s \\
\hline
7 & Reduce and remove build dependencies & 638 MiB & 145 MiB & 70.3 s \\
\hline
8 & Change base to eclipse-temurin & 630 MiB & 8 MiB & 72.5 s \\
\hline
9 & Change base to groovy:jdk21-alpine & 689 MiB & +59 MiB & 72.4 s \\
\hline
10 & Change base to sapmachine & 612 MiB & 77 MiB & 68.9 s \\
\hline
11 & Clean npm cache & 581 MiB & 31 MiB & 76.1 s \\
\hline
12 & Use older java 17 version & 560 MiB & 21 MiB & 76.6 s \\
\hline
13 & Remove sonarscanner & 536 MiB & 24 MiB & 76.7 s \\
\hline
14 & Use java 21 for extended lifetime & 559 MiB & +23 MiB & 78.4 s \\
\hline
\end{tabular}
\caption{Steps to reduce the container size and their impact.}
\label{table:dockermini}
\end{table*}

As detailed in \Cref{table:dockermini}, an extensive optimization effort reduced the image size from 1.4 GiB to 559 MiB, representing a 60\% reduction. As a result, the overall build time was also significantly decreased. Techniques included switching from Debian to Alpine as the base image, removing build dependencies, purging caches, and excluding obsolete tools. Each change was carefully evaluated for both performance and maintainability.  

\subsection{Running in GitHub Actions}\label{sec:runactions}

The final stage of the migration was the deployment of the refactored pipeline on GitHub Actions. This served as both a compliance requirement and a proof-of-concept for cross-platform portability.  

The migration process followed three phases: (1) establishing two separate repositories (one for COBOL projects, one for pipeline code), (2) converting the original \texttt{jenkinsfile} into YAML workflows, and (3) adapting the abstraction layer to interface with GitHub Actions’ API.  

Separation of repositories preserved architectural clarity and enabled role-based access control: COBOL developers were limited to application code, while DevOps engineers maintained pipeline logic. GitHub mechanisms such as \texttt{CODEOWNERS}, repository restrictions, and pre-receive hooks enforced this separation.  

The workflow itself was structured in two parts. The first configured the execution environment: specifying the container image, authenticating to the registry, and injecting secrets through GitHub’s secure storage. The second defined the pipeline steps: checking out repositories, marking safe directories, executing Groovy scripts, and performing cleanup in case of failure.  

The abstraction layer was updated to retrieve repository metadata from GitHub’s environment variables, ensuring that platform-specific details remained isolated. Functions were standardized to read, process, and return environment information consistently.  

Finally, execution was managed by GitHub runners, which provided containerized environments to run the pipeline. This architecture ensured resource isolation, efficient scaling, and compliance with security requirements.

\section{Evaluation}
\label{sec:evaluation}

This section first presents the evaluation of the migration including unit and integration testing and performance benchmarks to then present a qualitative product review with relevant stakeholders.

\subsection{Functional evaluation}

The experimental evaluation was designed to verify the functionality and efficiency of the pipeline. Three categories of tests were conducted: pipeline tests, container tests, and performance benchmarking.  

\subsubsection*{Pipeline tests}
To validate the functional correctness of the pipeline we used the \emph{Spock} testing framework, i.e., a behaviour-driven development tool for Java/Groovy integrated into the Gradle build system. Test specifications for core pipeline components were written in Spock’s declarative syntax, making use of \verb|when-then| and \verb|expect| constructs to define preconditions, actions, and assertions. Gradle orchestrated the execution of these tests during the build lifecycle, compiling pipeline code and dynamically executing Spock.  

Each \verb|when-then| test consisted of an action under test (e.g., invoking a method with specific parameters) followed by assertions confirming that the outcome matched expectations. For scenarios where stimulus and verification could be unified, the \verb|expect| construct provided a concise alternative.  

The unit test suite was integrated into a GitHub Actions workflow, ensuring continuous validation of code changes on every repository push. Beyond synthetic test cases, end-to-end validation was performed on a dedicated secondary repository containing five production-representative COBOL source files with their associated unit tests and build configurations. This enabled comprehensive integration testing, covering COBOL code processing, test suite execution, and overall workflow reliability.

All unit tests executed through Gradle and Spock passed successfully when compilation succeeded. Build failures were occasionally observed due to syntax errors or unsupported Jenkins DSL functions, which served as valuable early indicators of quality issues requiring refactoring. Integration testing on the dedicated COBOL repository confirmed that the pipeline could process real COBOL code, execute associated test suites, and maintain end-to-end workflow integrity.

However, trial executions also highlighted critical challenges, including the need to explicitly close mainframe connections after failures, discrepancies in expected versus actual container filesystem structures, and unintended modifications of shell commands. Addressing these issues led to targeted improvements in robustness. 

\subsubsection*{Container tests}
To guarantee reliability of the runtime environment, a GitHub Actions workflow automated the container image build process. All core dependencies (viz., Java, Groovy, GnuCOBOL, COBOL Check, and Zowe CLI) were tested for successful installation by executing version commands. Build failure was enforced if any test failed. A post-build Groovy script confirmed compatibility of the image with the pipeline’s runtime environment, including library access and script execution.  
This design ensured that only correctly configured images could progress through the pipeline.

Build-phase verification confirmed correct installation and execution of all dependencies, while runtime validation confirmed Groovy functionality and containerized script execution. Importantly, these tests also validated that functionality was preserved during the container image optimization.

\subsubsection*{Performance benchmarking}
To quantify performance improvements achieved through the migration, execution times of the \emph{original Jenkins pipeline} and the \emph{refactored GitHub Actions pipeline} were compared.  

Identical workloads were used in both settings, consisting of COBOL modules and unit tests.\footnote{Note that in this context the parallelism of the CI/CD pipeline was not a relevant factor, as the original pipeline is inherently sequential due to the serialized nature of mainframe interactions.} Execution times were measured using native platform metrics, with manual stopwatch verification for accuracy. Each configuration was executed five times to minimize the impact of runtime variability. The analysis focused on differences in mean runtime and variability between the two pipelines.  
Performance benchmarking demonstrated a dramatic reduction in execution times. The Jenkins-based pipeline averaged 724 seconds with a standard deviation of 47.3 seconds. By contrast, the GitHub Actions pipeline achieved a mean runtime of 130 seconds with a deviation of 9.5 seconds. This corresponds to an 82.1\% reduction in execution time.  

We stress that this comparison includes both architectural changes and platform differences; therefore, the measured improvement should be interpreted as an upper bound rather than an isolated effect of the refactoring.
While platform and infrastructure differences likely account for much of this improvement, the magnitude of change underscores the advantages of the migrated pipeline. A controlled experiment running the refactored pipeline on Jenkins could further isolate the effects of refactoring versus platform choice, though organisational access constraints prevented this test.

Please note that while the evaluation demonstrates clear improvements, 
it was not possible to validate portability across multiple platforms beyond GitHub Actions, 
due to organizational policies.  
Additionally, continuous deployment scenarios could not be tested in production.  
These limitations should be considered when generalizing results.

\subsection{Product Review}

A formal product review was conducted with Bankdata’s pipeline migration team to assess readiness for production deployment. Feedback was generally positive, particularly regarding the preservation of functionality during migration to GitHub Actions. However, several areas for improvement were identified.  

\subsubsection*{User experience}
Stakeholders noted that the GitHub Actions interface provided limited visual granularity compared to Jenkins. Whereas Jenkins allows fine-grained “step-level” feedback for each pipeline stage, the migrated pipeline appears as a single integrated workflow. Reviewers suggested introducing additional modularization to make it easier to localize failures. However, the project deliberately retained a monolithic script to maximize maintainability and portability across platforms. This reflects a conscious trade-off: improving error visibility for developers versus maintaining tool-agnostic design principles.  

Another important observation concerned developer familiarity with GitHub Actions. Since many COBOL developers are accustomed to mainframe workflows and rely on a single integrated development environment, the switch to GitHub’s web interface was perceived as disruptive. As a potential solution, the review highlighted the usefulness of the GitHub Actions extension for Visual Studio Code, which integrates workflow monitoring, build logs, and secret management directly into the IDE. This approach allows developers to remain within a familiar environment, smoothing the transition.  

\subsubsection*{Technology choices}
The choice of Groovy as the pipeline language was discussed at length. While YAML is usually the native format for pipeline specifications, Groovy was selected to ensure continuity with the Jenkins implementation and to preserve cross-platform independence. This decision was understood by stakeholders, but some raised concerns that Groovy might introduce a higher barrier to entry for developers less familiar with the language. Documentation and training were identified as necessary measures to address this gap.  

Security and compliance were also a central topic. The management of secrets within GitHub Actions was scrutinized, particularly in relation to the risk of inadvertent exposure through logs or artifacts. Reviewers recommended strengthening safeguards, including more restrictive access policies, additional log filtering, and regular audits of secret usage. Artifact storage strategies were likewise discussed, with agreement that organizational policies should guide whether GitHub’s native artifact storage is sufficient or whether integration with external systems is required.  

\subsubsection*{Cultural and organizational aspects}
Beyond technical feedback, the review underscored the importance of cultural alignment. The migration team emphasized that COBOL developers must be included in the adoption process to ensure that their workflows are preserved or improved. This echoes for example the following recommendation:

\begin{quote}
``Collaborate and communicate: This is a culture-focused play \textellipsis{} 
it is essential to include the mainframe development teams when adopting DevOps across the organization, 
even if the mainframe teams may still utilize different tools and process. \textellipsis{} 
doing so will result in a significant overall return on investment of the DevOps transformation.'' 
\cite{devopsadap}
\end{quote}

By combining technical validation with stakeholder feedback, the product review confirmed that the pipeline was functionally sound and ready for further adoption, while also identifying key areas—visualization, training, and security—for continued improvement. The balance struck between portability and usability will guide the next iteration of the migration strategy.

\section{Discussion}
\label{sec:discussion}

\subsection{Lessons Learned}

The migration process yielded a range of technical, organizational, and security insights that can inform similar initiatives in comparable contexts. These lessons were not only technical in nature, but also highlighted the importance of organizational alignment and security awareness in the design and operation of critical pipelines.  

\subsubsection{Technical Lessons}

From a technical perspective, one of the most significant insights was the importance of isolating platform-specific logic into a dedicated abstraction layer. This architectural choice is critical for achieving long-term portability, as it ensured that the overall logic of the pipeline remained intact even when migrating from Jenkins to GitHub Actions. By encapsulating platform dependencies within a well-defined boundary, the migration effort avoided widespread refactoring. In addition, containerization emerged as a central strategy for improving both performance and reproducibility. By embedding all required dependencies into a pre-built container image, the pipeline no longer relied on runtime installation of tools and libraries, which had previously introduced delays and potential points of failure. This shift improved execution times, but also provided a consistent environment across different runs, thereby reinforcing reproducibility and compliance.  

Note that while the custom abstraction layer and shell functions enabled decoupling from Jenkins, they also increase the importance of maintaining a well-defined and stable internal library, which now underpins core pipeline functionality and must be carefully governed to avoid technical debt. In addition, these abstractions may reduce the level of support available from AI-based development assistants, which have extensive knowledge of Jenkins semantics but may have  difficulty inferring the behaviour of custom functions.

Finally, while reducing vendor lock-in was a central goal of the migration, we acknowledge that the use of GitHub Actions does not in itself provide full cross-platform portability. Although the abstraction layer mitigates most platform-specific dependencies, achieving complete CI/CD agnosticism remains an open challenge.

\subsubsection{Organizational Lessons}

Equally important were the organizational lessons learned throughout the project. One recurring theme was the necessity of involving COBOL development teams early in the process. Their participation is invaluable in preventing resistance to change and ensuring that the new pipeline design aligned with their established workflows. Without their engagement, the transition to GitHub Actions could have introduced friction or even rejection, but their input helped ensure that the new solution was perceived as an enabler rather than a disruption.  

Another organizational insight concerned the trade-offs between modularity and usability. While a modular design allows for reusability and flexibility, it can also create steep learning curves for developers who are not familiar with the structure. In this case, maintaining monolithic workflow scripts within GitHub Actions simplified onboarding and provided a clear entry point for new users, even though it came at the expense of reduced modularity compared to a highly decoupled design. The experience thus underscored the importance of balancing architectural purity with practical usability, especially in environments where teams with diverse technical backgrounds must collaborate on a shared system.  

\subsubsection{Security Lessons}

Security considerations also played a crucial role in shaping the pipeline. One important lesson was that masking secrets in logs is not sufficient in itself. Although GitHub Actions provides mechanisms to hide sensitive values, these safeguards do not eliminate the possibility of accidental disclosure through misconfigurations or malicious exploitation of logs. A more robust approach requires dedicated mechanisms to prevent the extraction of secrets entirely, such as strict access controls, careful validation of workflows, and continuous monitoring.  

Another security-related lesson was that dependencies cannot be evaluated solely in terms of minimization. While reducing unnecessary third-party components is generally desirable, some dependencies play a vital role in ensuring stability and long-term maintainability. For example, retaining tools that contribute to reproducibility, even if they slightly increase the size of the container image, may ultimately strengthen the overall security posture by reducing variability and unpredictability in execution environments. This finding emphasizes that security and stability should be considered together when evaluating dependencies, rather than treating them as conflicting goals.

Although the case study focuses on COBOL systems, the architectural principles—namely abstraction of platform-specific logic and containerized dependency management—are independent of programming language. These patterns are applicable to other legacy or modern stacks, particularly where pipelines exhibit tight coupling and runtime dependency installation.

\subsection{Threats to Validity}

Despite the positive results, several threats to validity must be acknowledged in order to properly contextualize the findings of this work.  

First, the observed performance improvements may be subject to \textit{performance bias}. While the migration led to an 82.1\% reduction in runtime, part of this improvement could stem from differences in the underlying infrastructure rather than architectural enhancements alone. Jenkins and GitHub Actions operate on distinct hardware and platform configurations, which may have influenced the results. Without running the refactored pipeline under identical conditions on Jenkins, it is not possible to attribute the performance gains exclusively to the architectural changes.  

A second concern relates to \textit{generalizability}. Although the abstraction layer and containerization strategies aimed to be tool-agnostic, the implementation was validated only within GitHub Actions. Other CI/CD platforms, such as GitLab or CircleCI, were not tested, which means that the conclusions drawn from this study cannot be generalized to all CI/CD ecosystems without further empirical evidence.  

The scope of the evaluation also introduces limitations. Due to organizational policies, it was not possible to validate continuous deployment scenarios. The evaluation was thus confined to continuous integration, and as a result, the findings do not necessarily extend to the broader class of CI/CD processes. Since deployment introduces unique challenges such as rollback management and production monitoring, future work would be required to verify that the proposed design scales effectively to continuous deployment contexts.  

Finally, the possibility of \textit{researcher bias} must be acknowledged. A subset of the authors of this work were themselves part of the migration team, which raises the risk of unintentional bias in both evaluation and reporting. While efforts were made to mitigate this—such as relying on objective benchmarking and incorporating stakeholder feedback—complete impartiality cannot be guaranteed. Independent replication by external teams would therefore provide valuable confirmation of the validity of these findings.

\section{Related Work and Conclusions}
\label{sec:conclusions}

There are comparatively few peer-reviewed, industrial case studies that dissect end-to-end software delivery pipelines operating in production environments.

As far as COBOL modernization and transformation efforts are concerned, the majority of academic and industrial research has concentrated on approaches to either transform legacy programs into modern languages or cope with the inherent complexity of large, mission-critical codebases. This orientation reflects the enduring dependence of banks, governments, and enterprises on COBOL systems, coupled with the ongoing shortage of developers fluent in the language.
For example, De Marco et al.~\cite{DBLP:conf/icsm/MarcoIA18} report on the migration of a newspaper delivery system from COBOL to Java. This case study highlights the challenges of preserving functional equivalence during a large-scale migration effort, while demonstrating how testing frameworks and iterative verification are crucial to sustaining business continuity. Gaudl and Brune~\cite{DBLP:conf/webist/GaudlB22} propose a pattern-based approach for transforming COBOL applications into RESTful web services. The emphasis on service wrapping and exposure via standard protocols resonates with our pipeline’s goal of reducing dependency on specific tools while supporting incremental modernization.
Melo et al.~\cite{melo2025human} examine human--AI collaboration in the modernization of COBOL-based systems in the public sector. Their case study on the Department of Government Efficiency (DOGE) illustrates the sociotechnical challenges of integrating AI support into critical modernization processes, including issues of knowledge transfer, organizational resistance, and cybersecurity risks.
Similarly, Hans et al.~\cite{DBLP:conf/sigsoft/Hans0YOKSSMKS25}, Gandhi et al.~\cite{DBLP:conf/llm4code/Gandhi0KVM24}, and Deknop et al.~\cite{DBLP:conf/vissoft/DeknopMBFZ21} all focus on enhancing the transformation and maintainability of COBOL systems, whether by ensuring semantic equivalence in COBOL-to-Java translations, leveraging large language models for code refinement, or introducing visualization techniques to support effective refactoring.  

As far as DevOps and CI/CD are concerned, much of the literature consists of handbooks and conceptual guidance, while detailed reports from industry settings—covering architecture, migration constraints, operational trade-offs, and measured outcomes—are rarer. Among the notable exceptions is~\cite{DBLP:conf/sigsoft/DebroyMB18}, which documents how containerization and DevOps principles were applied to build lean CI/CD pipelines and quantitatively improve build performance. Zampletti et al.~\cite{DBLP:conf/icsm/ZampettiGBP21} study CI/CD pipelines as evolving software artifacts, showing that pipeline configurations undergo continuous restructuring to address maintainability, performance, and technological changes. Rostami Mazrae et al. \cite{DBLP:journals/ese/MazraeMGD23} further highlight that migration between CI/CD platforms is a common practice in industry, identifying key drivers, trade-offs, and barriers through a qualitative study of practitioners. The majority of the people interviewed about migrations from Travis to GitHub Actions, but a pair also involve Jenkins to GitLab or GitHub Actions, like in our case. Chopra et at. \cite{DBLP:conf/icsm/ChopraG25} study the evolution of multi-CI service adoption, showing that nearly one in five projects on GitHub use multiple CI services, and that migration between services is common.
On the security side, Moyón et al.~\cite{DBLP:conf/profes/MoyonSPMB20} present an industry case study on integrating security standards directly into DevOps pipelines (DevSecOps), highlighting process adaptations and compliance impacts. In parallel, Cox~\cite{supply_chain} synthesized five decades of supply-chain security concerns and trends in software reuse, providing contemporary motivation for hardening build and delivery pipelines. 

In this paper we presented an industrial case study on the migration of Bankdata’s 
legacy Jenkins pipeline to a modernized, containerized architecture. 
The migration resulted in a more portable design that is not tied to Jenkins, 
a runtime improvement of approximately {82\%}, reduced complexity through 
repository refactoring and dependency management, and improved maintainability and auditability. These achievements demonstrate that even highly 
regulated organizations can successfully modernize their CI/CD processes without 
compromising compliance or stability.  
Beyond the immediate 
context, the project offers a transferable blueprint for other organizations in 
regulated industries that wish to modernize legacy systems while safeguarding 
security and compliance.  

Several directions for future work remain open. Further validation 
is required across multiple CI/CD platforms beyond GitHub Actions in order to 
fully assess the generalizability of the tool-agnostic design. Another promising 
avenue is the replacement of the Zowe CLI with a dedicated Java SDK to 
provide tighter integration, stronger type safety, and a cleaner interface for 
pipeline operations. The development of reusable workflow templates could also 
accelerate adoption across additional projects by standardizing common tasks while 
preserving necessary flexibility. Finally, the expansion of automated test coverage 
would enable continuous deployment scenarios, which could not be explored within 
this study due to organizational constraints. Extending the approach to deployment 
would shorten release cycles and improve resilience by ensuring earlier 
and more reliable detection of failures.

% ***EXPANDEDBIBSTYLE: \bibliographystyle{IEEEtran}

% ***EXPANDEDBIBFILE: \bibliography{biblio}
% ****EXPORTBEGS: \bibliography{main.bbl}

% ****EXPORTENDS: \bibliography{main.bbl}

\end{document}